# Evaluation of the Fourth Millennium Development Goal Realisation Using Robust and Nonparametric Tools Offered by Data Depth Concept

**Ewa Kosiorowska[c], Daniel Kosiorowski[a], Zygmunt Zawadzki[b]**
[a]*Cracow University of Economics, Department of Statistics, Poland*
[b]*Cracow University of Economics, Master Studies, Poland*
[c]*General Practitioner, Internist, Poland*

**Abstract**: We briefly communicate results of nonparametric and robust evaluation of effects of *the Fourth Millennium Development Goal of United Nations*. Main aim of the goal was reducing by two thirds, between 1990-2015, the under five months child mortality. Our novel analysis was conducted by means of very powerful and user friendly tools offered by the *Data Depth Concept* being a collection of multivariate techniques basing on multivariate generalizations of quantiles, ranges and order statistics. Results of our analysis are more convincing than results obtained using classical statistical tools.

## 1. Introduction

The sample mean vector (MV) and the sample covariance matrix (COV) have been the standard estimators of location and scatter in the multivariate statistics. They are *working horses* of the classical comparative analysis being a basis for a statistical evaluation of various socio-economic goals. Unfortunately, economic data sets very often contain *outliers* or *inliers* of a various kind, what makes the MV and the CLS useless due to their extreme sensitivity to atypical observations. Even one point departing from the main part of the data may destroy the estimation results obtained using them. In a *static*, comparative Economics, we very often cannot use powerful tools of the classical multivariate statistics basing on the mean vector, the covariance matrix and the normality assumptions. In the *dynamic* Economy the situation is even more complicated.

This paper aims at presenting opportunities of conducting effective empirical researches in robust and nonparametric manner by means of not well known but very powerful and user friendly tools offered by the so called *Data Depth Concept*. The tools are implemented in our free R package *DepthProc* (see Kosiorowski and Zawadzki, 2014), which is available on CRAN servers. Our main motivations relate to our attempts of finding an empirical dimension of recent developments in *Theory of Cooperative Dynamic Games* (CDG) (see Petrosjan and Yeung, 2012). Our interest in the CDG originates from our research program concerning robust analysis of economic data streams. We have found that under certain circumstances a group of cooperating economic agents can be treated as biological system processing data streams (see Kosiorowski, 2014).

Let us only recall that *Robust Statistics* aims at identifying a tendency represented by an influential majority of data and detecting observations departing from that tendency (see Marona et al., 2006). Main ideas of the robust statistics are closely tied with a notion of an influential majority of agents, investors, consumers - ruling the behaviour of a certain economic system. Unfortunately, methods of the robust statistics are rather rarely present in the current economic discussions, and in empirical or theoretical studies.

We understand robustness of the estimator in terms of the influence function (IF) and the finite sample breakdown point (BP) – for further details see Maronna et. al. (2006). Let us recall that for a given distribution $F$ in $\mathbb{R}^d$ and an $\varepsilon > 0$, the version of $F$ contaminated by an $\varepsilon$ amount of an arbitrary distribution $G$ in $\mathbb{R}^d$ is denoted by $F(\varepsilon, G) = (1-\varepsilon)F + \varepsilon G$. The **influence function** (IF) of an estimator $T$ at a given $\mathbf{x} \in \mathbb{R}^d$ for a given $F$ is defined

$$IF(\mathbf{x}; T, F) = \lim_{\varepsilon \to 0^+} \left( T(F(\varepsilon, \delta_x)) - T(F) \right) / \varepsilon, \qquad (4)$$

where $\delta_{\mathbf{x}}$ is the point-mass probability measure at $\mathbf{x} \in \mathbb{R}^d$.

The $IF(\mathbf{x}; T, F)$ describes the relative effect (influence) on $T$ of an infinitesimal point-mass contamination at $\mathbf{x}$, and measures the local robustness of $T$. An estimator with bounded IF (with respect to a given norm) is therefore robust (locally, as well as globally) and very desirable.

Let $\mathbf{X}^n = \{\mathbf{X}_1, ..., \mathbf{X}_n\}$ be a sample of size $n$ from $\mathbf{X}$ in $\mathbb{R}^d$, $d \geq 1$. The **replacement breakdown point** (BP) of a scatter estimator $V$ at $\mathbf{X}^n$ is defined as

$$BP(V, \mathbf{X}^n) = \left\{ \frac{m}{n} : tr\left( V(X^n) V(X_m^n)^{-1} + V(X^n)^{-1} V(X_m^n) \right) = \infty \right\}, \qquad (5)$$

where $\mathbf{X}_m^n$ is a contaminated sample resulting from replacing $m$ points of $\mathbf{X}^n$ with arbitrary values, $tr()$ denotes trace of the matrix. The BP point serves as a measure of global robustness, while the IF function captures the local robustness of estimators.

In a context of empirical studies of the CDG, we considered data sets related to *the Millennium Declaration*. Let us only briefly recall, that at the beginning of the century, world leaders come together at the *United Nations* and agreed on a general vision for the future through *the Millennium Declaration*. The Millennium Development Goals (MDGs) express the principles of human dignity, equality, and free the world from extreme poverty (see UN Report, 2014). Although, the MDGs can be successfully evaluated for example within the SDEA (see Stone, 2002) framework - a success of these considerations crucially depends on a preliminary statistical analysis of the issue. In this work, we focus our attention on a statistical layer of the MDGs achievement evaluation.

For showing usefulness of the Data Depth Concept (DDC) tools, let us consider an evaluation of results of a program known as *The Fourth Millennium Development Goal of United Nations* (**4MG**). Main aim of the goal was reducing by two thirds, between 1990-2015, the under five months child mortality. Although, all MDGs are in fact only a general declaration of good-will, in a context of the 4MG, the World Health Organization promotes four main strategies for obtaining the goal: **1.** appropriate home care and timely treatment of complications for new-borns; **2.** integrated management of childhood illness for all children under five years old; **3.** expanded programme of immunisation; **4.** infant and young child feeding. It should be stressed that in developing countries most deaths in the first year of life happen immediately after delivery or in the first days. Unfortunately medical care for the new-born children is not available. Obstetric intervention protected from asphyxia, infections, treatment of childbirth complications is commonly neglected. One can improve infant mortality among children under five years old by dissemination of vaccinations for childhood diseases, antibiotics for bacterial infections and oral rehydration therapy for diarrhoea (see Goldenberg and Jobe, 2001, and references therein). Taking on the concrete means by countries for obtaining the 4MG is beyond a direct control of the United Nations.

The rest of the paper is organized as follows: in Section 2, basic notions related to the Data Depth Concept and further used techniques are briefly described. In Section 3, the empirical analysis of the 4MG conducted by means of the depth tools is presented. The paper ends with some conclusions and references. All data sets and examples considered within the paper are available after installing the *DepthProc* R package.

## 2. Selected data depth statistical tools

### 2.1 General notions of the DDC

**Data depth concept** (**DDC**) was originally introduced as a way to generalize the concepts of median and quantiles to the multivariate framework. A depth function $D(\cdot, F)$ associates with any $\mathbf{x} \in \mathbb{R}^d$ a measure $D(\mathbf{x}, F) \in [0,1]$ of its centrality with respect to (w.r.t.) a probability measure $F \in \mathcal{P}$ over $\mathbb{R}^d$ or w.r.t. an empirical measure $F_n \in \mathcal{P}$ calculated from a sample $\mathbf{X}^n$.

The larger the depth (closer to one) of $\mathbf{x}$, the more central $\mathbf{x}$ is w.r.t. to $F$ or $F_n$. A point (or a set of points) for which a depth function attains a maximum is called **a multivariate median** induced by this depth. In a context of economic application of the DDC, it is worth focusing of our attention on **the weighted $L^p$ depth**. The weighted $L^p$ depth from a sample $\mathbf{X}^n = \{\mathbf{x}_1, ..., \mathbf{x}_n\} \subset \mathbb{R}^d$ is computed as

$$D(\mathbf{x}; \mathbf{X}^n) = \frac{1}{1 + \frac{1}{n}\sum_{i=1}^{n} w\left(\|\mathbf{x} - \mathbf{X}_i\|_p\right)}, \qquad (1)$$

where $w$ is appropriate non-decreasing and continuous on $[0, \infty)$ weight function, and $\|\cdot\|_p$ stands for the $L^p$ norm (when $p = 2$ we have the usual Euclidean norm).

The weighted $L^p$ depth function *in a point*, has the low breakdown point (BP) and unbounded influence function IF (the BP - expresses a minimal fraction of "bad" points in a sample making a procedure useless, the (IF) measures sensitivity of a procedure on a point contamination for the BP and IF formal definitions). On the other hand, the weighted $L^p$ depth induced medians (multivariate location estimators) are globally robust with the highest BP for any reasonable estimator. The weighted $L^p$ medians are also locally robust with bounded influence functions for suitable weight functions. For the theoretical properties see Zuo (2004).

**A symmetric projection depth** $D(x, X)$ of a point $x \in \mathbb{R}^d$, $d \geq 1$ is defined as

$$D(x, X)_{PRO} = \left[1 + sup_{\|u\|=1} \frac{\left|u^T x - Med\left(u^T X\right)\right|}{MAD\left(u^T X\right)}\right]^{-1}, \qquad (2)$$

where Med denotes the univariate median, $MAD(Z) = Med(|Z - Med(Z)|)$. Its sample version denoted by $D(x, X^n)$ or $D(x, F_n)$ is obtained by replacing population $F$ by its empirical counterpart $F_n$ calculated from the sample $X^n$. This depth is affine invariant and $D(x, F_n)$ converges uniformly and strongly to $D(x, F)$. The affine invariance ensures that our proposed inference methods are coordinate-free, and the convergence of $D(x, X^n)$ to $D(x, X)$ allows us to approximate $D(x, F)$ by $D(x, X^n)$ when $F$ is unknown. Induced by this depth, multivariate location and scatter estimators have very high breakdown points and bounded Hampel's influence function (for further details see Zuo, 2003). Procedures induced by L$^p$ depth are robust but not very robust in opposition to procedures induced by the projection depth.

The set of points for which depth takes value not smaller than $\alpha \in [0,1]$ is multivariate analogue of the quantile and is called the $\alpha-$ central region,

$$D_\alpha(\mathbf{X}^n) = \{\mathbf{x} \in \mathbb{R}^d : D(\mathbf{x}, \mathbf{X}) \geq \alpha\}. \tag{3}$$

Every depth function induces a *classification rule* in a form: classify a point to a population in which the point has a maximal value. In an opposition to the density function, the depth function has a global nature i.e., e.g., that it expresses a centrality of a point w.r.t. a whole sample. This property is its disadvantage in the context of classification of objects. A successful depth based classifier needs a *local* version of depth. A very successful concept of local depth was proposed in Paindaveine and Van Bever (2013). For a depth function $D(\cdot, P)$, the **local depth** with the locality parameter $\beta \in (0,1]$ and w.r.t. a point $x$ is the usual depth conditioned on a specially chosen neighbourhood of the point $x$. The neighbourhood covers $\beta$ fraction of the points concentrated around $x$.

Next, very useful for the economic applications example of depth, is **regression depth** introduced in Rousseeuw and Hubert (1999). The deepest regression estimator is an example of the *robust regression*. An up to date and detailed presentation of the DDC concept can be found in Mosler (2013). Below we only briefly present multivariate generalisation of well known **quantile-quantile plot**, a generalization of **Wilcoxon rang sum statistic** and **the scale curve** being a multivariate nonparametric scatter functional.

**2.2 Nonparametric and robust comparisons of multivariate samples**

**The Depth vs. Depth** plots (**DD-plots**)}, introduced by Liu et al. (1999), is an user friendly two-dimensional graph allowing us for a visual comparison of two samples of any dimension (multivariate quantile-quantile plot).

For two samples $\mathbf{X}^n = \{X_1, ..., X_n\}$ from $F$, and $\mathbf{Y}^m = \{Y_1, ..., Y_m\}$ from $G$, DD-plot is defined

$$DD(F_n, G_m) = \left\{\left(D(\mathbf{z}, F_n), D(\mathbf{z}, G_m)\right), \mathbf{z} \in \left\{\mathbf{X}^n \cup \mathbf{Y}^m\right\}\right\}. \tag{4}$$

A difference in a location manifests in a form of a star-shaped pattern, whereas a difference in a scale manifests as a moon-shaped pattern.

It is easy to notice, that having two samples $\mathbf{X}^n$ and $\mathbf{Y}^m$ and using any depth function, we can compute depth values in a combined sample $\mathbf{Z}^{n+m} = \mathbf{X}^n \cup \mathbf{Y}^m$, assuming the empirical distribution calculated basing on all observations, or only on observations belonging to one of the samples $\mathbf{X}^n$ or $\mathbf{Y}^m$. If we observe $X_i$'s depths are more likely to cluster tightly around the center of the combined sample, while $Y_l's$ $Y_{l'}s$ depths are more likely to scatter outlying positions, then we conclude $\mathbf{Y}^m$ was drawn from a distribution with larger scale.

The **depth based multivariate Wilcoxon rang sum test** is especially useful for the multivariate scale changes detection. For the samples $\mathbf{X}^m = \{\mathbf{X}_1,...,\mathbf{X}_m\}$, $\mathbf{Y}^n = \{\mathbf{Y}_1,...,\mathbf{Y}_n\}$, and a combined sample $\mathbf{Z} = \mathbf{X}^n \cup \mathbf{Y}^m$ the **Wilcoxon statistic** is defined as

$$S = \sum_{i=1}^{m} R_i, \qquad (5)$$

where $R_i$ denotes the rang of the i-th observation, $i = 1,...,m$ in the combined sample $R(\mathbf{x}_l) = \#\{\mathbf{z}_j \in \mathbf{Z} : D(\mathbf{z}_j, \mathbf{Z}) \leq D(\mathbf{x}_l, \mathbf{Z})\}, l = 1,...,m.$ The distribution of $S$ is symmetric about $E(S) = 1/2 m(m+n+1)$, its variance is $D^2(S) = 1/12\, mn(m+n+1)$. For theoretical properties of the test see Li and Liu (2004) and Zuo and He (2006).

For sample depth function $D(x, Z^n)$, $x \in \mathbb{R}^d$, $d \geq 2$, $Z^n = \{z_1,...,z_n\} \subset \mathbb{R}^d$, $D_\alpha(Z^n)$ denoting $\alpha-$ central region, we can define **the scale curve**

$$SC(\alpha) = \left(\alpha, vol(D_\alpha(Z^n))\right) \subset \mathbb{R}^2, \qquad (6)$$

for $\alpha \in [0,1]$.

The scale curve shows an expansion of a volume of $\alpha-$ central region (the expansion around a multivariate median) in a relation to the outlyingness index $\alpha$ (see Mosler, 2013, and references therein for further details).

## 3. Empirical Studies

In our study we jointly considered following variables being officially recommended by the United Nations and having a medical justification (Goldberger and Jobe, 2001):

1. *Children under 5 months mortality rate per 1,000 live births* (**Y₁**)

2. *Infant mortality rate (0-1 year) per 1,000 live births* (**Y₂**)

3. *Children 1 year old immunized against measles, percentage* (**Y₃**)

Data sets were obtained from http://mdgs.un.org/unsd/mdg/Data.aspx and are available within our free package *DepthProc*:

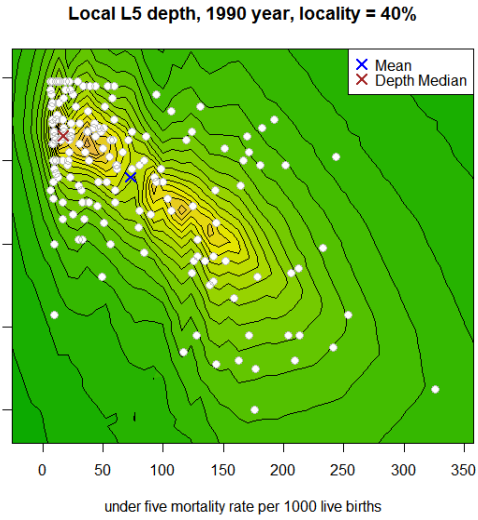
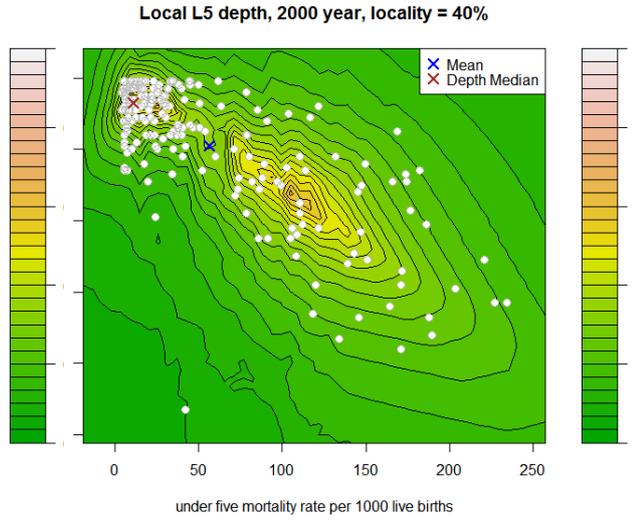

**Fig.1.** The 4MG in 1990: $L^5$ depth contour plot $Y_1$ vs. $Y_3$, calculations DepthProc.

**Fig.2.** The 4MG in 2000: $L^5$ depth contour plot $Y_1$ vs. $Y_3$, calculations DepthProc.

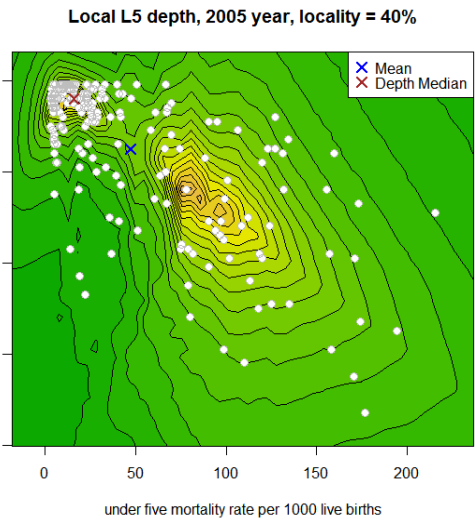
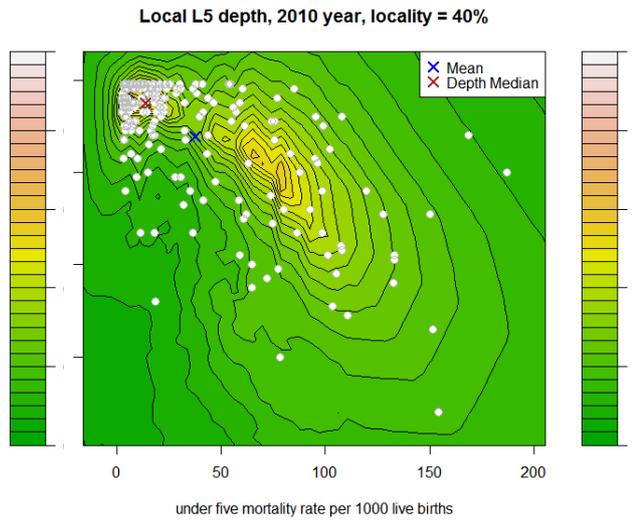

**Fig.3.** The 4MG in 2005: $L^5$ depth contour plot $Y_1$ vs. $Y_3$, calculations DepthProc.

**Fig.4.** The 4MG in 2010: $L^5$ depth contour plot $Y_1$ vs. $Y_3$, calculations DepthProc.

Fig. 1–4 show weighted $L^5$ depth contourplots with locality parameter $\beta = 0.4$ for countries in 1990, 2000, 2005, 2010 considered w.r.t. variables $Y_1$ and $Y_3$, whereas Fig. $5-8$ present weighted $L^5$ depth contourplots with locality parameter $\beta = 0.4$ for countries in 1990, 2000, 2005, 2010 w.r.t. variables $Y_2$ and $Y_3$. Although we can notice a socio-economic development between 1990 and 2011 - the clusters of developed and developing countries are still evident in 2011 as they were in 1990.

For assessing changes in location of the centers and scatters of the data, between 1990 and 2011, we calculated $L^1$ **medians** (a robust measure of a center) and $L^5$ **weighted covariance matrices** for $(Y_1, Y_2, Y_3)$ (a robust measure of a scatter) which are presented in Tab. 1

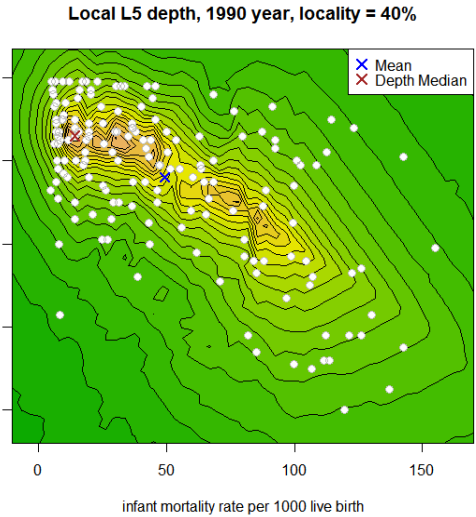

**Fig.5.** The 4MG in 1990: $L^5$ depth contour plot $Y_2$ vs. $Y_3$, calculations DepthProc.

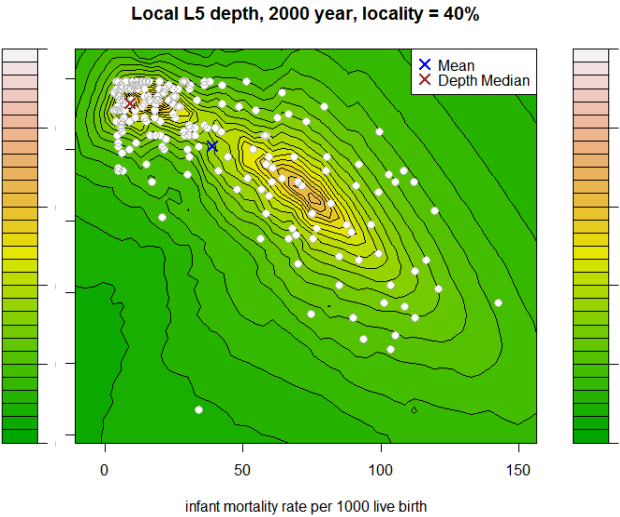

**Fig.6.** The 4MG in 2000: $L^5$ depth contour plot $Y_2$ vs. $Y_3$, calculations DepthProc.

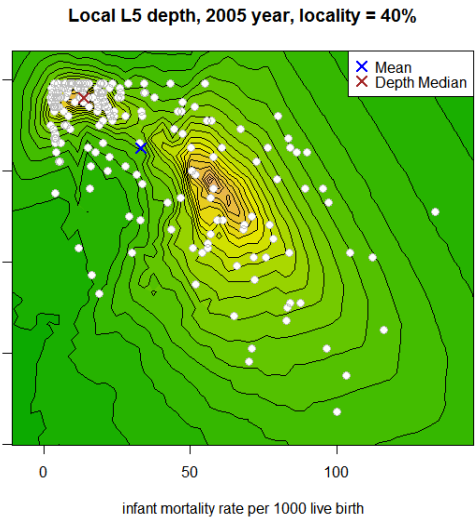

**Fig. 7:** The 4MG in 2005: $L^5$ depth contour plot $Y_2$ vs. $Y_3$, calculations DepthProc.

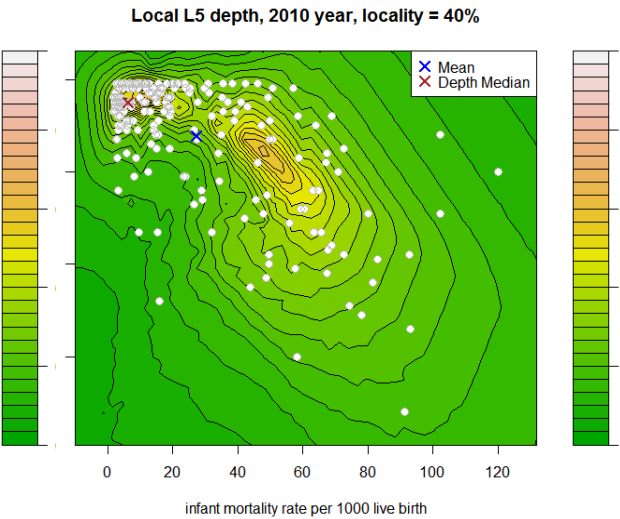

**Fig. 8:** The 4MG in 2010: $L^5$ depth contour plot $Y_2$ vs. $Y_3$, calculations DepthProc.

It is easy to notice a significant development in absolute values for a majority of the countries, both in the location as well as in the scale of the phenomenon. Fig. 10 and 12. present DD-plots for inspecting location changes between 1990 and 2011 between 2000 and 2011 for countries considered w.r.t. all variables $(Y_1, Y_2, Y_3)$ and Fig. 9 and 11 present DD-plots for inspecting scale changes for the same data. Patterns presented on these plots may be treated as evident arguments for positive changes between 1990 and 2011. Please note that the considered data suggest non-normality and an existence of outliers and hence the classical methods of samples comparison are not applicable here.

**Tab. 1:** Estimated location characteristics for 4MG indicators.

|  | $L^1$ median | | | Projection median | | | Mean vector | | |
|---|---|---|---|---|---|---|---|---|---|
| year | $Y_1$ | $Y_2$ | $Y_3$ | $Y_1$ | $Y_2$ | $Y_3$ | $Y_1$ | $Y_2$ | $Y_3$ |
| **1990** | 49.7 | 41.9 | 83.0 | 23.2 | 19.6 | 86.0 | 72.85 | 49.28 | 76.06 |
| **1995** | 41.3 | 35.8 | 85.0 | 26.4 | 22.2 | 90.0 | 62.67 | 42.97 | 79.04 |
| **2000** | 30.4 | 25.3 | 89.0 | 17.2 | 14.3 | 93.0 | 56.19 | 38.76 | 81.09 |
| **2005** | 23.9 | 20.3 | 91.0 | 18.4 | 15.5 | 94.0 | 47.05 | 33.10 | 84.97 |
| **2010** | 18.8 | 17.3 | 93.0 | 13.8 | 11.7 | 95.0 | 37.47 | 27.25 | 87.74 |

**Source:** Our own calculations, DepthProc package.

$$COV_{L^5}(1990) = \begin{pmatrix} 2133 & 1295 & -397 \\ 1295 & 817 & -240 \\ -397 & -240 & 264 \end{pmatrix}; COV_{L^5}(2000) = \begin{pmatrix} 1449 & 910 & -384 \\ 910 & 594 & -240 \\ -384 & -240 & 251 \end{pmatrix}$$

$$COV_{L^5}(2005) = \begin{pmatrix} 1107 & 705 & -297 \\ 705 & 468 & -192 \\ -297 & -192 & 218 \end{pmatrix}; COV_{L^5}(2010) = \begin{pmatrix} 745 & 497 & -180 \\ 497 & 342 & -121 \\ -180 & -121 & 175 \end{pmatrix}$$

We performed induced by $L^2$ depth multivariate *Wilcoxon* test for a scale change detection for the variables ($Y_1$, $Y_2$, $Y_3$) in 1990 and in 2011 and obtained: **W=17538** and **p-value=0.0363**. We can conclude therefore, that both the scale and the location, changed in this period. Fig. 13. presents scale curves for the countries considered in the period 1990--2011 jointly w.r.t. all variables. The three-dimensional scatters around the three-dimensional $L^2$ medians decreased significantly in this period. Fig. 14-16 present Student depth contour plots for the variables $Y_1$, $Y_2$ and $Y_3$ considered separately in 1990-2011. The Student depth relates to one dimensional location-scale problem, i.e, joint estimation of a location and scale parameters. *The Student median* is an effective and robust alternative for a mean and standard deviation. It indicates a value between the median and the mode. Its contour plot can be treated as very powerful method of inspecting distributional assumptions. Further details can be found in Mizera and Müller (2004). Fig. 14 – 16 show positive tendencies as to changes of location and scale of variables together with retaining distributional properties - asymmetry related to clusters of rich and poor countries remains nearly on the same level.

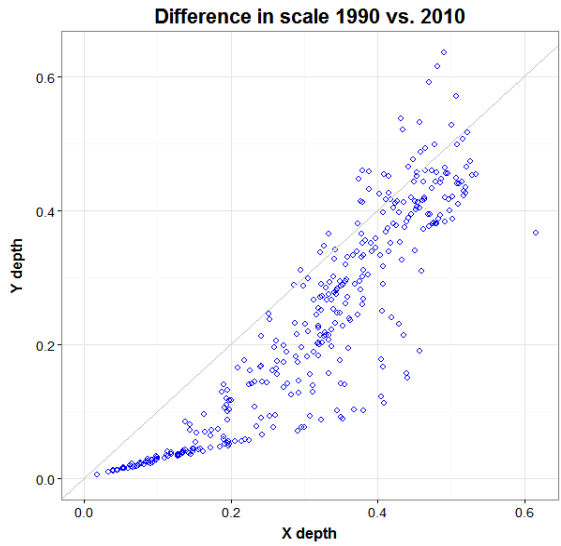
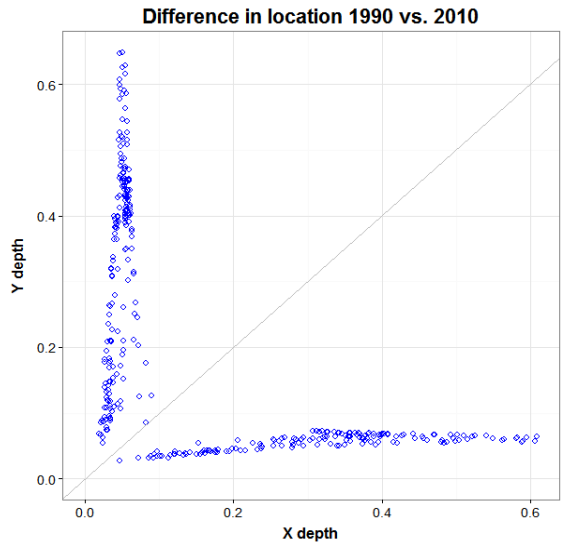

**Fig.9:** DD plot for inspecting scale differences for $(Y_1, Y_2, Y_3)$ in 1990 and 2010.

**Fig.10:** DD plot for inspecting location differences for $(Y_1, Y_2, Y_3)$ in 1990 and 2010.

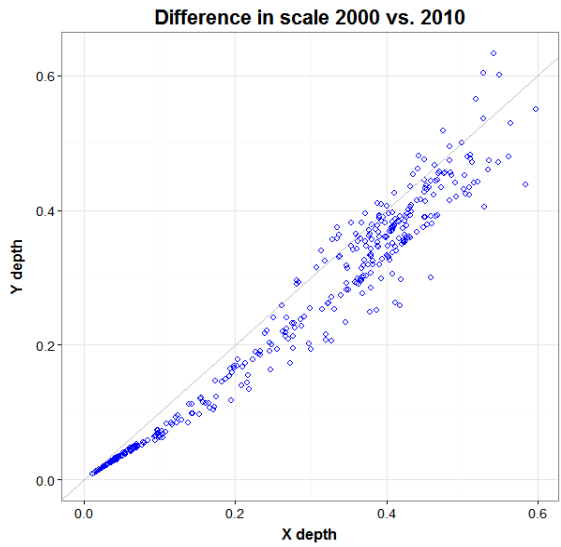
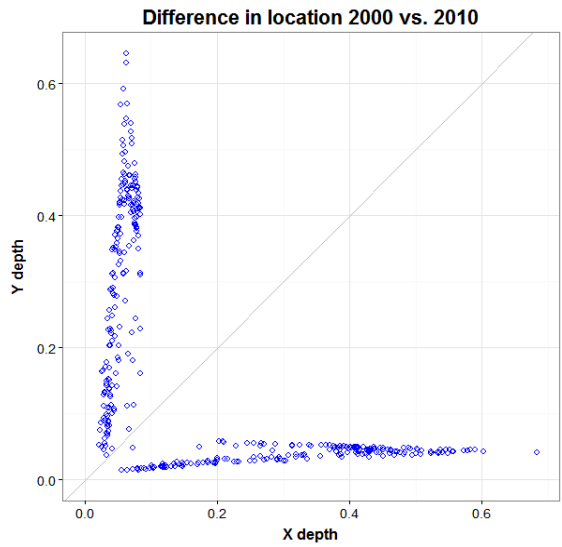

**Fig.11:** DD plot for inspecting scale differences for $(Y_1, Y_2, Y_3)$ in 2000 and 2010.

**Fig.12:** DD plot for inspecting location differences for $(Y_1, Y_2, Y_3)$ in 2000 and 2010.

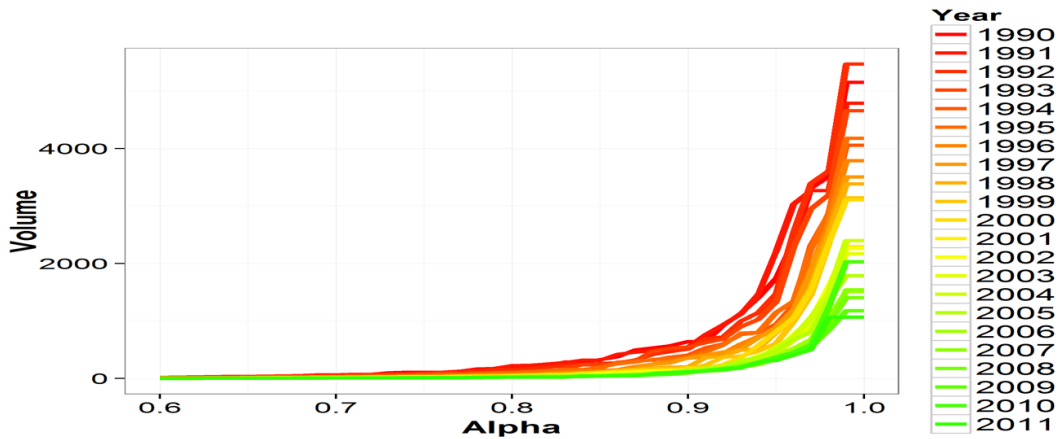

**Fig.13.** Scale curves for $(Y_1, Y_2, Y_3)$ 1990-2011, calculations DepthProc.

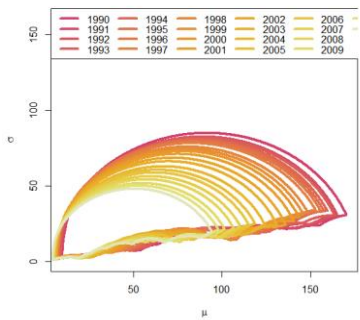 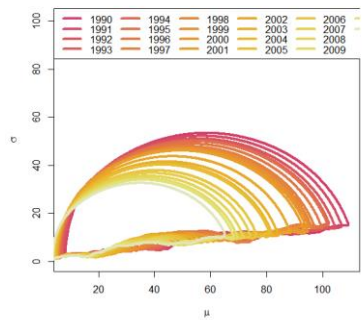 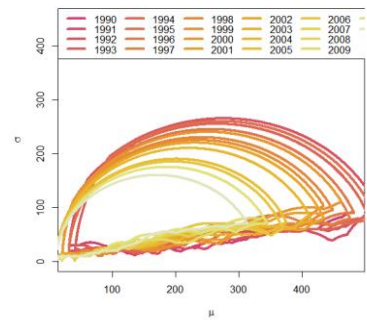

**Fig.14.** Student depth contour plots – $Y_1$ in 1990—2011, calculations DepthProc.

**Fig.15.** Student depth contour plots – $Y_2$ in 1990—2011, calculations DepthProc.

**Fig.16.** Student depth contour plots – $Y_3$ in 1990—2011, calculations DepthProc.

Fig. 17 presents a relation between variables $Y_1$ and $Y_2$ obtained by means of the simple deepest regression (DR - a robust method) and the popular least squares regression (LS - a nonrobust method) in 1990 year. Fig. 18 presents an analogous situation in 2011 year. Please note, that the strong relation between these variables is evident only when using the robust regression method (the slopes in 1990: LS -0.28; DR -1.36, and the slopes in 2011: LS -0.34; DR -1.067).

**The results of the analysis lead us to following conclusions**:

1. There are big chances for obtaining the 4MG. In 2010 the decrease in the under five months child mortality was about 40% with robust estimates used.
2. For the variables considered jointly, both multivariate as well as univariate scatters decreased in 1990-2011. The decrease is especially evident in a closeness to the three-dimensional medians.
3. A comparison of the Student medians in 1990--2011 indicates significant one-dimensional tendencies for obtaining the 4MG. Asymmetries between rich and poor countries remain evident.
4. The estimated simple deepest regressions are arguments for a correct choice of the 4DG goal realisation indicators. Relations between variables (interpreted as inputs and outputs) are evident.

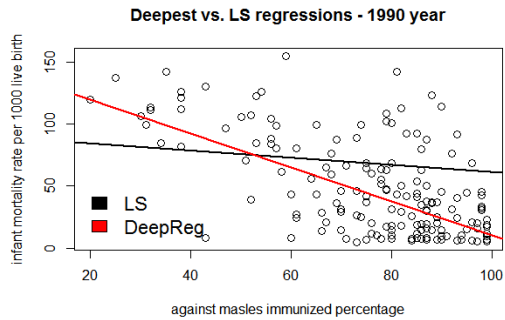 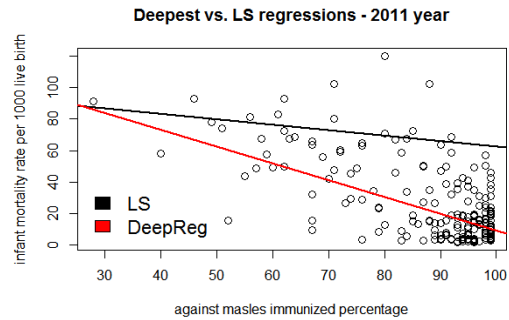

**Fig. 17.** Deepest vs. Least Squares regressions, $Y_1$ vs. $Y_3$ in 1990, calculations DepthProc.

**Fig.18.** Deepest vs. Least Squares regressions, $Y_1$ vs. $Y_3$ in 2011, calculations DepthProc.

## 4. Conclussions

In the paper we presented a statistical evaluation of the 4MD conducted by means of robust and nonparametric tools offered by the DDC. The analysis was performed using data offered by the United Nations, which departure from normality and contain outliers. Results of the analysis are in favour for big chances for achieving the 4MD in a near future. The differences between *poor* and *rich* countries still remain evident however. The DDC offers a comprehensive family of statistical tools for effects of various social programs evaluation purposes.

**Acknowledgement**

Daniel Kosiorowski thanks for the NCS financial support DEC-011/03/B/HS4/01138.